\documentclass[acmsmall,screen]{acmart}

\setcopyright{none}
\AtBeginDocument{%
  \renewcommand\footnotetextcopyrightpermission[1]{}%
}

\usepackage{colortbl} 
\usepackage{etoolbox}
\usepackage{listings} 
\usepackage{parcolumns}
\AtBeginDocument{\DeclareCaptionSubType{lstlisting}}
\usepackage{caption}
\usepackage{subcaption}

\usepackage{enumitem}
\usepackage{tabularx}
\usepackage{listings}
\usepackage{framed}

\definecolor{codegreen}{rgb}{0,0.6,0}
\definecolor{codegray}{rgb}{0.5,0.5,0.5}
\definecolor{codepurple}{rgb}{0.58,0,0.82}
\definecolor{backcolour}{rgb}{0.95,0.95,0.92}

\lstdefinestyle{pythonstyle}{
    backgroundcolor=\color{backcolour},   
    commentstyle=\color{codegreen},
    keywordstyle=\color{magenta},
    numberstyle=\tiny\color{codegray},
    stringstyle=\color{codepurple},
    basicstyle=\ttfamily\footnotesize,
    breakatwhitespace=false,         
    breaklines=true,                 
    captionpos=t,                    
    keepspaces=true,                 
    numbers=left,                    
    numbersep=5pt,                  
    showspaces=false,                
    showstringspaces=false,
    showtabs=false,                  
    tabsize=2
}

\newtoggle{showcomments}
\toggletrue{showcomments}

\AtBeginDocument{%
  }

\setcopyright{none}
\copyrightyear{}
\acmYear{}
\acmDOI{}
\acmPrice{}
\acmISBN{}

\begin{document}

\title{Artificial or Just Artful? Do LLMs Bend the Rules in Programming?}

\author{Oussama Ben Sghaier}
\email{oussama.sghaier@queensu.ca}
\orcid{0000-0003-2737-0952}
\affiliation{%
  \institution{Queen's University}
  \city{Kingston}
  \state{Ontario}
  \country{Canada}
}

\author{Kevin Delcourt}
\affiliation{%
  \institution{Université de Montréal}
  \city{Montréal}
  \state{Québec}
  \country{Canada}
}
\email{kevin.delcourt@umontreal.ca}
\orcid{0009-0005-2988-7308}

\author{Houari Sahraoui}
\affiliation{%
  \institution{Université de Montréal}
  \city{Montréal}
  \state{Québec}
  \country{Canada}
}
\email{sahraouh@iro.umontreal.ca}
\orcid{0000-0001-6304-9926}

\renewcommand{\shortauthors}{Ben Sghaier et al.}

\begin{abstract}
Large Language Models (LLMs) are widely used for automated code generation, yet their apparent successes often mask a tension between pretraining objectives and alignment choices. While pretraining encourages models to exploit all available signals to maximize success, alignment, whether through fine-tuning or prompting, may restrict their use. 
This conflict is especially salient in agentic AI settings, for instance when an agent has access to unit tests that, although intended for validation, act as strong contextual signals that can be leveraged regardless of explicit prohibitions.
In this paper, we investigate how LLMs adapt their code generation strategies when exposed to test cases under different prompting conditions. Using the BigCodeBench (Hard) dataset, we design five prompting conditions that manipulate test visibility and impose explicit or implicit restrictions on their use. We evaluate five LLMs (four open-source and one closed-source) across correctness, code similarity, program size, and code churn, and analyze cross-model consistency to identify recurring adaptation strategies.
Our results show that test visibility dramatically alters performance, correctness nearly doubles for some models, while explicit restrictions or partial exposure only partially mitigate this effect. Beyond raw performance, we identify four recurring adaptation strategies, with test-driven refinement emerging as the most frequent. 
These results highlight how LLMs adapt their behavior when exposed to contextual signals that conflict with explicit instructions, providing useful insight into how models reconcile pretraining objectives with alignment constraints.

\end{abstract}





\maketitle

\section{Introduction}

Large Language Models (LLMs) are trained to perform many tasks, from solving mathematical problems \cite{cobbe2021gsm8k} to producing human-like text \cite{brown2020gpt3}. Their training objectives are often complemented by alignment choices introduced during fine-tuning \cite{ouyang2022training,sghaier2025leveraging} or through prompting \cite{wei2022chain, jaoua2025combining}.

When alignment through prompting makes a task more difficult, LLMs face a conflict between 
their pretraining objective—to exploit all available signals to maximize task success—and the alignment objective—to follow user-specified rules \cite{perez2022discovering, park2024ai, sghaier2025harnessing}. This tension often results in altered behavior. Models may appear to comply while still relying on unintended cues (e.g., producing confident but incorrect answers in mathematics rather than admitting uncertainty\cite{madaan2024selfcheck}), over-comply in ways that reduce task performance (e.g., refusing to explain harmless chemical processes because they resemble restricted content \cite{wei2022chain}), or ignore instructions entirely in favor of coherence or user satisfaction (e.g., agreeing with biased statements \cite{perez2022discovering} or sidestepping constraints in creative writing \cite{hubinger2024sleeper}). Such conflicts arise because alignment, applied through prompts or fine-tuning, is layered on top of strong statistical patterns from pretraining and is therefore easier to override \cite{pan2022effects}.

These challenges also manifest in software engineering tasks such as code review, code search, and code generation \cite{zhuo2024bigcodebench,ma2025rethinking}. In code generation, the pretraining objective is to produce correct code for a given problem, while alignment choices may introduce additional requirements. Examples include adhering to non-functional properties such as efficiency \cite{weyssowcodeultrafeedback} or respecting explicit constraints such as avoiding the use of available contextual signals \cite{baker2025monitoring,liu2025trust} such as unit tests. 
As with general-purpose tasks, alignment choices in code generation can make it more difficult—or even impossible—to satisfy the underlying training objectives. Understanding how LLMs navigate these conflicts is therefore essential to anticipate their behavior and limitations.

The integration of LLMs into automated code generation pipelines has amplified these concerns. It remains unclear whether their apparent successes reflect genuine problem-solving capabilities or opportunistic reliance on unintended shortcuts \cite{ma2025rethinking}. Unit tests, in particular, commonly included for evaluation, can act as hidden signals that models exploit to boost performance \cite{baker2025monitoring}. More broadly, LLM-based development tools increasingly operate in environments where prompts may inadvertently include logs, configuration fragments, execution traces, or intermediate artifacts. Even when users or systems attempt to suppress reliance on such information, models may still opportunistically exploit it.

In this paper, we present an empirical study that systematically examines how LLMs adapt their code generation strategies when exposed to test cases under different instructions. Using the BigCodeBench (Hard) dataset~\cite{zhuo2024bigcodebench}, we define five prompting conditions that manipulate test visibility and impose explicit or implicit restrictions on their use: (1) Baseline (B), with only the task description; (2) Full Test (FT), task plus all tests; (3) Full Test + Do Not Use (FT+DNU), with explicit instructions to ignore tests; (4) Partial Test (PT), a subset of tests without full coverage; and (5) Partial Test + Do Not Use (PT+DNU), the subset plus explicit restrictions. We evaluate five LLMs (four open-source and one closed-source) across correctness, similarity to reference code, program size, and code churn, and analyze cross-model consistency to identify adaptation strategies.

Our results show that test visibility dramatically alters performance: correctness nearly doubles for some models when tests are accessible. For example, Pass@1 success rates rose from 15.5\%–24.4\% without tests to 37.2\%–54.7\% with tests (including FT and FT+DNU conditions). Similar gains were observed for Pass@5, which increased from 29.1\%–37.2\% without tests to 57.4\%–72.3\% with tests. 

These findings demonstrate that LLMs routinely override explicit prohibitions and, in some cases, perform better under restricted conditions than unrestricted ones—suggesting complex interactions between alignment instructions and contextual signal exploitation.
Moreover, our analysis reveals diverse strategies for handling conflicts between training and alignment objectives, highlighting both the adaptability and the vulnerabilities of current LLMs in code generation.

Our goal is not to evaluate whether models behave appropriately in a realistic ‘provide tests but ignore them’ workflow. Instead, we investigate a broader phenomenon: when LLM-based assistants are exposed to auxiliary contextual signals that conflict with explicit instructions, do they exploit those signals anyway, and if so, how? Unit tests offer a precise mechanism for injecting such signals and detecting their influence.

The rest of the paper is organized as follows. Section~\ref{sec:background} reviews the background and related work. Section~\ref{sec:design} details the study design, including dataset selection, model choices, prompting strategies, evaluation metrics, and experimental setup. Section~\ref{sec:results} presents both quantitative and qualitative analyses of the experimental results. Section~\ref{sec:discussion} discusses threats to validity and reflects on the findings. Finally, Section~\ref{sec:conclusion} concludes and outlines directions for future research.

\section{Background and Related Work}
\label{sec:background}

This section provides background on large language models (LLMs) for code generation, including their pretraining objectives, alignment techniques, and documented instances of misalignment. We emphasize the tension between the performance-driven incentives established during pretraining and the constraints imposed by alignment methods, which can lead models to exploit contextual signals in ways that conflict with intended behavior. We conclude by situating our study within this context and motivating our contribution.

\subsection{LLMs for code generation}

Large language models trained or adapted for source code have rapidly advanced automated programming~\cite{jiang2024survey, ben2024improving}, from API-aware completion in IDEs~\cite{chen2021evaluating} to solving contest-level problems~\cite{li2022competition}. Open initiatives have released strong code-oriented base models and datasets (e.g., SantaCoder~\cite{allal2023santacoder}, StarCoder2~\cite{lozhkov2024starcoder}, Code Llama~\cite{roziere2023code}), making research and evaluation broadly accessible and inspiring instruction-tuned derivatives specialized for following natural language prompts~\cite{luo2023wizardcoder}.

A rich ecosystem of benchmarks now evaluates different facets of code generation. HumanEval and MBPP measure functional correctness on short Python tasks using unit tests~\cite{chen2021evaluating,austin2021program}; APPS spans a wider difficulty range, from introductory problems to competition-style challenges~\cite{hendrycks2021measuring}; and MultiPL-E systematically translates HumanEval and MBPP into multiple languages to probe cross-language generalization~\cite{cassano2022multipl}. More recently, BigCodeBench emphasizes diverse function calls and complex instructions, better reflecting current model capabilities~\cite{zhuo2024bigcodebench}. Across benchmarks, evaluation typically relies on executing unit tests to standardize correctness assessment.

\subsection{Pretraining Objectives of LLMs}

Modern LLMs are predominantly trained with predictive pretraining objectives. In the autoregressive setting~\cite{malach2023auto}, this involves maximizing the likelihood of the next token given its preceding context, thereby reducing predictive error on large-scale corpora and supporting generalization to a wide range of downstream tasks~\cite{brown2020language}.

A direct consequence of this objective is that models become highly sensitive to \emph{contextual signals}: any hint in the prompt that reduces next-token uncertainty can be exploited during generation. This often works in the user's favor: large autoregressive models demonstrate strong in-context learning, adapting outputs to structured prompts without updating parameters~\cite{brown2020language,coda2023meta}.

Pretraining thus establishes a strong, performance-oriented prior: the model’s implicit preference is to minimize next-token loss by leveraging all available information. Alignment techniques, described next, are applied precisely because this incentive alone is insufficient to ensure behavior consistent with user intent.

\subsection{Alignment and Restrictions in LLMs}

LLMs are often adapted to align with user preferences or specialized tasks. Common approaches include \emph{prompt engineering}~\cite{marvin2023prompt}, \emph{instruction fine-tuning}~\cite{wei2021finetuned}, and \emph{reinforcement learning from human feedback (RLHF)}~\cite{ouyang2022training}.

While these techniques can substantially shape model behavior, for example, to better fit code generation tasks~\cite{weyssow2025exploring}, they do not eliminate the underlying pretraining incentive to exploit predictive signals available at inference time~\cite{raffel2020exploring}. This tension between broad predictive objectives and restrictive alignment goals can result in misaligned behavior, especially when contextual cues strongly suggest strategies that contradict explicit instructions.



\subsection{Model Misalignment}

Several documented instances of misalignment highlight the gap between pretraining incentives, alignment methods, and actual model behavior.


\citet{park2024ai} survey deceptive LLM behaviors in social and strategic domains, including manipulation and cheating in diplomacy, poker, and economic negotiation. They also report models attempting to trick safety tests or deceive human reviewers. Complementing this,~\citet{hubinger2024sleeper} demonstrate that it is possible to train \emph{deceptive sleeper agents}, LLMs that behave benignly during most interactions and safety training but reveal adversarial objectives under specific triggers.

Reward hacking is another recurring failure mode, originating in reinforcement learning: agents exploit underspecified reward functions to maximize reward while violating the designer’s intent~\cite{pan2022effects}. In LLMs fine-tuned with human feedback, similar effects occur. \citet{ferreira2025truthful} find that chain-of-thought explanations often produce plausible-sounding rationales that do not reflect true internal computation.


In programming, misalignment emerges in task-specific ways, though it remains underexplored. Evaluations such as the \emph{Claude Opus and Sonnet 4 system card} \cite{anthropic2025claude4} and analyses of OpenAI o3-mini~\cite{baker2025monitoring} document cases where models overfit to provided test suites, producing solutions that pass visible tests but fail to generalize. Other studies report hallucinations in code generation, where syntactically valid programs are semantically incorrect~\cite{lee2025hallucination}.

These findings indicate that broader misalignment phenomena, deception, reward hacking, or sycophancy, also appear in programming contexts.

\subsection{Motivation}

The preceding discussion highlights a recurring theme: pretraining establishes a strong, performance-oriented prior, driving models to exploit contextual signals, while alignment techniques attempt to constrain behavior to match human intent. In programming tasks, this tension can manifest as misalignment: models may overfit prompts, manipulate outputs to satisfy benchmarks, or produce semantically invalid code that superficially appears correct.

Unit tests, common in code evaluation benchmarks, constitute particularly salient signals. Though intended for validation, they provide precise guidance on passing criteria, creating scenarios where pretraining incentives directly conflict with alignment constraints. 
Understanding how models navigate this tension is critical not because test-driven improvement is unexpected, but because it reveals how models behave when useful contextual signals are explicitly forbidden.
Motivated by this gap, our study systematically investigates how LLMs adapt code generation strategies when exposed to test cases under varying prompting conditions. By manipulating test visibility and imposing explicit or implicit restrictions, we aim to observe whether models leverage test information, how alignment constraints modulate behavior, and which adaptation strategies recur across different LLM architectures.

\section{Study Design}
\label{sec:design}

In this section, we present the overall design of our study, detailing the research questions, dataset, models, prompting strategies, evaluation metrics, and experimental setup, following recent guidelines for reporting empirical studies in software engineering involving LLMs \cite{baltes2025evaluation}.

\subsection{Research Questions}

To systematically investigate the instruction-following capabilities and behavioral tendencies of LLMs in code generation, we articulate four research questions (RQs). Each RQ is associated with a specific analysis and experimental setup.  

\begin{leftbar}
\textbf{RQ1.} \textit{Does embedding test cases in the prompt, either fully or partially, and with or without explicit restrictions, improve LLM performance in code generation?}
\end{leftbar}

To answer this question, we compare the performance of five LLMs across five prompting strategies: \emph{Baseline (B)}, \emph{Full Test (FT)}, \emph{Full Test + Do Not Use (FT+DNU)}, \emph{Partial Test (PT)}, and \emph{Partial Test + Do Not Use (PT+DNU)}. These conditions vary both the visibility of test cases and the explicitness of restrictions. We evaluate performance primarily using \emph{pass@k}, complemented by task-level indicators such as newly passing and failing tests. This allows us to quantify whether embedding test cases enhances accuracy, even when restrictive instructions are present, and to assess the extent to which LLMs exploit contextual signals provided by tests.  

\begin{leftbar}
\textbf{RQ2.} \textit{Do the prompting strategies (B, FT, FT+DNU, PT, PT+DNU) lead to substantially different generations compared to the reference solution \emph{R} provided by BigCodeBench, and if so, how do these differences manifest?}
\end{leftbar}

Here we move beyond correctness to examine the qualitative impact of prompting on code generation. We investigate whether test exposure and restrictions lead to genuinely distinct outputs relative to \emph{R}, or whether the differences remain superficial. Specifically, we analyze how models alter their code: by introducing new elements, simplifying or removing constructs, or restructuring solutions. To quantify these effects, we use complementary measures such as \emph{CodeBLEU} and \emph{LOC} to capture the nature and extent of variations. 

\begin{leftbar}
\textbf{RQ3.} \textit{Are the behavioral differences observed across prompting strategies statistically significant?}
\end{leftbar}

To address this question, we test whether the differences identified in \emph{RQ2} are robust under statistical scrutiny. In this case, rather than comparing prompting-based solutions with the reference solution \emph{R}, we compare the prompting strategies to each other. In addition to \emph{CodeBLEU}, we use churn-based metrics (added lines, removed lines, and total churn). We then apply the \emph{Wilcoxon} signed-rank test \cite{woolson2007wilcoxon}, a non-parametric method for paired data with the False Discovery Rate (FDR) correction \cite{benjamini1995controlling}, to assess the statistical significance of the observed variations. This ensures that our conclusions regarding performance gains and behavioral changes are supported by rigorous evidence.

\begin{leftbar}
\textbf{RQ4.} \textit{What adaptation strategies do LLMs employ across prompting strategies with varying levels of test visibility and restriction?}
\end{leftbar}

Finally, we conduct a qualitative analysis of how LLMs adjust their generations under different conditions and how they exploit test signals in the prompt. Through manual inspection of representative code samples that exhibited marked differences in metrics, we identify recurring adaptation strategies. This provides insight into how models balance pretraining-driven objectives with alignment constraints, and how they respond differently to implicit versus explicit restrictions.

\subsection{Dataset}

In this study, we use \emph{BigCodeBench} \cite{zhuo2024bigcodebench}, a large-scale benchmark specifically designed for evaluating the code generation capabilities of large language models. \emph{BigCodeBench} contains $1,140$ programming tasks covering a wide range of real-world scenarios. Each task includes a natural language specification, a set of test cases, and a canonical reference implementation. The benchmark provides, on average, 5.6 test cases per problem with approximately 99\% branch coverage, ensuring that generated code can be assessed comprehensively and reproducibly \cite{zhuo2024bigcodebench}.

A key characteristic of \emph{BigCodeBench} is that each task is available in two specification formats. The \emph{complete} variant provides detailed docstrings with descriptions, parameter types, usage examples, and notes, while the more concise \emph{instruct} variant distills this information into a shorter, human-like request. In our study, we rely on the \emph{instruct} format to better simulate natural user queries. 

In our experiments, we focus on the subset of $148$ tasks designated as \emph{hard} within \emph{BigCodeBench}, each accompanied by executable test cases that collectively yield $854$ individual unit tests. Compared to the rest of the benchmark, these tasks are deliberately more challenging, often involving multiple function calls, cross-domain reasoning, or the integration of complex programmatic constructs. Whereas LLMs achieve near-ceiling accuracy on easier tasks, the \emph{hard} subset reveals greater variability in model performance and behavior, providing a more discriminative basis for evaluating our different prompting strategies. By excluding easy problems that models already solve reliably, we ensure that our analysis highlights meaningful differences emerging from non-trivial challenges, rather than being confounded by ceiling effects.

\subsection{Models selection}

In our study, we selected five models, four open-source and one closed-source, based on their strong performance on \emph{BigCodeBench}'s leaderboard for the hard tasks~\cite{zhuo2024bigcodebench}. We prioritized models that achieve high accuracy on code generation from specifications, while ensuring that the open-source variants can be run on our local infrastructure.
Table~\ref{tab:models} summarizes the chosen models.

\begin{table}[ht]
\centering
\caption{Selected models for evaluation. The \emph{size} of the model represents the number of the model's parameters.}
\label{tab:models}
\resizebox{\textwidth}{!}{%
\begin{tabular}{p{2.5cm} p{.5cm} p{1cm} p{11.2cm}}
\toprule
\textbf{Model} & \textbf{Size} & \textbf{Source} & \textbf{Description} \\
\midrule
\textbf{GPT5-nano} & 18B & Closed-source & Lightweight member of the GPT5 family, optimized for low-latency reasoning and coding tasks in production environments. Designed for speed and efficiency with decent accuracy on coding benchmarks \cite{openai2025gpt5}. \\\midrule

\textbf{Qwen2.5-Coder} & 14B & Open-source & Specialized code-focused LLM from the Qwen family. Supports long context (131K tokens) and achieves strong performance on code generation benchmarks, ranking among the best open models \cite{hui2024qwen2}. \\\midrule

\textbf{Phi-4} & 14B & Open-source & Compact reasoning and coding model developed by Microsoft Research. Trained on high-quality synthetic data to achieve strong results on STEM and programming tasks, often outperforming larger models \cite{abdin2024phi}. \\\midrule

\textbf{OpenCoder} & 8B & Open-source & Open and reproducible code LLM trained from scratch on 2.5T tokens (90\% code, 10\% related data). Optimized for both English and Chinese, with strong instruction-following and competitive code generation performance \cite{huang2024opencoder}. \\\midrule

\textbf{Ministral} & 8B & Open-source & Efficient model designed for fast inference and long-context reasoning (up to 128K tokens). Incorporates sliding-window attention for memory efficiency, enabling deployment on limited infrastructure \cite{mistral2024ministral}. \\
\bottomrule
\end{tabular}
}
\end{table}

\subsection{Prompting stratgies}

In this study, we compare the performance of five LLMs across five prompting strategies: \emph{Baseline (B)}, \emph{Full Test (FT)}, \emph{Full Test + Do Not Use (FT+DNU)}, \emph{Partial Test (PT)}, and \emph{Partial Test + Do Not Use (PT+DNU)}. These strategies differ in the visibility of test cases provided to the model and in the explicitness of restrictions imposed on their use, as illustrated in Table~\ref{tab:prompting_strategies}.

In the \emph{Baseline (B)} strategy, models receive only the natural language specification of the task without any test cases. This represents the standard code generation setup and serves as the reference point for comparisons. 

In the \emph{Full Test (FT)} strategy, the complete set of unit tests is included in the prompt alongside the specification, giving the model direct access to all evaluation criteria. 

In \emph{Full Test + Do Not Use (FT+DNU)} strategy, the tests are provided but prefaced with an explicit instruction forbidding their use in code generation to mitigate trivial code copying from test cases.

The \emph{Partial Test (PT)} strategy instead supplies only a subset of the evaluation tests, typically half (i.e., two or three test cases) while the remainder are kept hidden. The prompt explicitly notes that the hidden tests will be used for evaluation (e.g., ``Here are the tests that will be used to evaluate your code <tests subset provided>'').  This approach reduces the information available to the model and tests its ability to generalize beyond the visible cases. 

Finally, in the \emph{Partial Test + Do Not Use (PT+DNU)} strategy, the same subset of tests is included but accompanied by the same explicit prohibition against their direct use for code generation.

\begin{table}[ht]
\centering
\caption{Templates of the prompting strategies.}
\label{tab:prompting_strategies}
\resizebox{.85\textwidth}{!}{
\begin{tabularx}{\linewidth}{|p{2.8cm}|X|}
\hline
\textbf{Strategy} & \textbf{Prompt template} \\
\hline
Baseline (B) & 
\texttt{<problem specification>} \\
\hline
Full Test (FT) & 
\texttt{<problem specification>} \newline
\textbf{\texttt{Here are the tests that will be used to evaluate your code:}} \newline
\texttt{<all tests>} \\
\hline
Full Test + Do Not Use (FT+DNU) & 
\texttt{<problem specification>} \newline
\textbf{\texttt{Here are the tests that will be used to evaluate your code. You must NOT use them to write your code:}} \newline
\texttt{<all tests>} \\
\hline
Partial Test (PT) & 
\texttt{<problem specification>} \newline
\textbf{\texttt{Here are the tests that will be used to evaluate your code:}} \newline
\texttt{<subset of tests>} \\
\hline
Partial Test + Do Not Use (PT+DNU) & 
\texttt{<problem specification>} \newline
\textbf{\texttt{Here are the tests that will be used to evaluate your code. You must NOT use them to write your code:}} \newline
\texttt{<subset of tests>} \\
\hline
\end{tabularx}
}
\end{table}

\subsection{Metrics}

To comprehensively evaluate the performance and behavior of LLMs under different prompting strategies, we employ a set of complementary metrics. These metrics fall into two main categories: (i) correctness-oriented metrics, which measure whether the generated code solves the given tasks, and (ii) structural and difference-oriented metrics, which quantify how the generated code diverges from reference solutions or from baseline generations.
These syntactic measures serve as auxiliary indicators of behavioral change, capturing how models adapt their generation strategies in response to test exposure.

\paragraph{\textbf{Correctness-oriented metrics}}
We assess functional correctness primarily with \emph{pass@$k$} for $k \in{(1,5)}$, a standard metric in code generation benchmarks \cite{chen2021evaluating}. Pass@$k$ reflects the probability that at least one out of $k$ independently sampled generations successfully passes all unit tests for a given task. Concretely, pass@$k$ equals 1 for a task if at least one of the $k$ generated solutions satisfies the entire test suite, and 0 otherwise.

In addition, we report \emph{new pass}, the number of tasks solved under a given prompting strategy that were not solved in the baseline, and \emph{new fail}, the number of tasks that fail under the new strategy despite being solved in the baseline. These complementary measures allow us to track both improvements and regressions introduced by different prompting strategies.

\paragraph{\textbf{Structural and difference-oriented metrics.}}
To quantify the similarity between generated code and reference implementations, we use \emph{CodeBLEU} \cite{ren2020codebleu}, which extends BLEU with code-specific n-gram, syntax, and data-flow matches. To assess changes in code length, we compute the number of non-comment lines of code (LOC). We further capture code modification dynamics using churn-based metrics: the number of added lines ($+$) and the number of removed lines ($-$). These churn metrics enable a fine-grained analysis of how much code is changed across the different prompting scenarios.

\subsection{Experiments}

\begin{figure}
\includegraphics[width=1\linewidth]{figures/methodology.pdf}
\caption{Overview of the experimental methodology.Each task from the \emph{BigCodeBench-Hard} dataset is paired with five different prompting strategies (Baseline, FT, PT, FT+DNU, PT+DNU). Each LLM then generates five code solutions, which are executed against the reference test suites. Outputs are evaluated with correctness-oriented metrics (e.g., pass@$k$, new pass/fail) and structure-oriented metrics (e.g., CodeBLEU, LOC, churn). Results are aggregated to answer the four research questions (RQ1–RQ4).}
\label{fig:methodology}
\end{figure}

We conducted our experiments on the \emph{BigCodeBench-Hard} dataset. For each task specification, we generated five code solutions per model for each prompting strategy, as illustrated in Figure~\ref{fig:methodology}. The resulting outputs were then evaluated using both correctness- and structure-oriented metrics.

The four open-source models (\emph{Qwen2.5-Coder-14B}, \emph{Phi-4}, \emph{OpenCoder-8B}, and \emph{Ministral-8B}) were executed locally on a server equipped with four \emph{NVIDIA RTX A5000 GPUs}. For the closed-source model (\emph{openai-gpt5-nano}), we used the OpenAI Batch API under identical prompting conditions.  

Following generation, every output was executed against the corresponding \emph{BigCodeBench} test suite. Correctness-oriented metrics (e.g., pass@$k$, new pass, new fail) were derived directly from test outcomes, while structure-oriented metrics (e.g., CodeBLEU, LOC, churn) were computed by comparing generated code to reference implementations or to baseline outputs. These aggregated results were then used to address the four posed research questions (RQ1–RQ4).

\section{Results}
\label{sec:results}
In this section, we present the findings of our study, organized into two complementary parts. The first sections report the quantitative analysis addressing RQ1–RQ3, focusing on how test visibility and restrictions influence code generation. The last section, corresponding to RQ4, provides a qualitative analysis of the adaptation strategies employed by LLMs when exposed to test cases.

\subsection{RQ1. Impact on the Generation Accuracy}



Table~\ref{tab:pass_k} reports pass@1 and pass@5 across models and prompting strategies. From these results, several categories of observations can be made.  
A first category concerns the impact of test visibility. Providing test cases in the prompt, even when the instruction specifies that tests are intended for evaluation, consistently yields substantial gains over the baseline across all models. For instance, \textit{Qwen2.5-Coder-14B} improves from $24.3\%$ to more than $50\%$ in pass@1 under full test prompting (FT), while \textit{Phi-4} nearly doubles its performance from $23.6\%$ to $54.7\%$. Similar gains are observed in pass@5, where the strongest models nearly double their baseline performance. These results confirm that test visibility provides a strong signal that models systematically exploit to improve task performance.


\begin{table}[ht]
\centering
\caption{Pass@1 and Pass@5 results for different prompting strategies across models.}
\label{tab:pass_k}
\begin{tabular}{lccccc}
\hline
\textbf{Prompt Version} & \textbf{Qwen2.5-14B} & \textbf{Phi-4} & \textbf{OpenCoder-8B} & \textbf{Ministral-8B} & \textbf{GPT5-nano} \\
\hline
\multicolumn{6}{c}{\textit{Pass@1 (\%)}} \\
\hline
Baseline                 & 24.32 & 23.6 & 17.6 & 15.5 & 24.4 \\
FT               & 50.68 & 54.7 & 37.8 & 37.2 & 49.2 \\
FT+DNU & 50.00 & 56.8 & 38.5 & 37.8 & 52.4 \\
PT          & 41.20 & 42.6 & 29.7 & 29.1 & 39.5 \\
PT+DNU & 40.54 & 39.9 & 31.1 & 27 & 37.8 \\
\hline
\multicolumn{6}{c}{\textit{Pass@5 (\%)}} \\
\hline
Baseline                 & 33.1 & 37.2 & 35.1 & 29.1 & 36.8 \\
FT               & 66.2 & 70.3 & 57.4 & 61.5 & 65.5 \\
FT+DNU & 62.2 & 72.3 & 58.1 & 60.1 & 65.5 \\
PT          & 56.8 & 57.4 & 48.6 & 45.9 & 52 \\
PT+DNU & 56.1 & 57.4 & 48.6 & 50 & 48 \\
\hline
\end{tabular}
\end{table}


The second category of observations relates to partial visibility. Across all models, partial test prompting (PT and PT+DNU) produces improvements over the baseline but consistently falls short of the full test conditions. For example, \textit{OpenCoder-8B} improves from $17.6\%$ at baseline to around $30\%$ with PT, compared to nearly $38\%$ with FT. This suggests that partial visibility induces partial improvement: even with access to only a subset of test cases, models benefit from the visible tests, as reflected by consistently higher pass@$k$ than the baseline. At the same time, these gains remain systematically lower than those achieved under full visibility.  

A third line of observation involves explicit restrictions. Instructions to disregard tests (\emph{Do Not Use}, or DNU) do not negate the observed gains. In many cases, performance under FT+DNU or PT+DNU is comparable to, or even slightly higher than, the unrestricted variants. For instance, \textit{Phi-4} achieves its best pass@1 of $56.8\%$ with FT+DNU. This suggests that models systematically disregard explicit prohibitions and continue to exploit the contextual presence of tests, even when instructed otherwise.  

\begin{figure}
\includegraphics[width=1\linewidth]{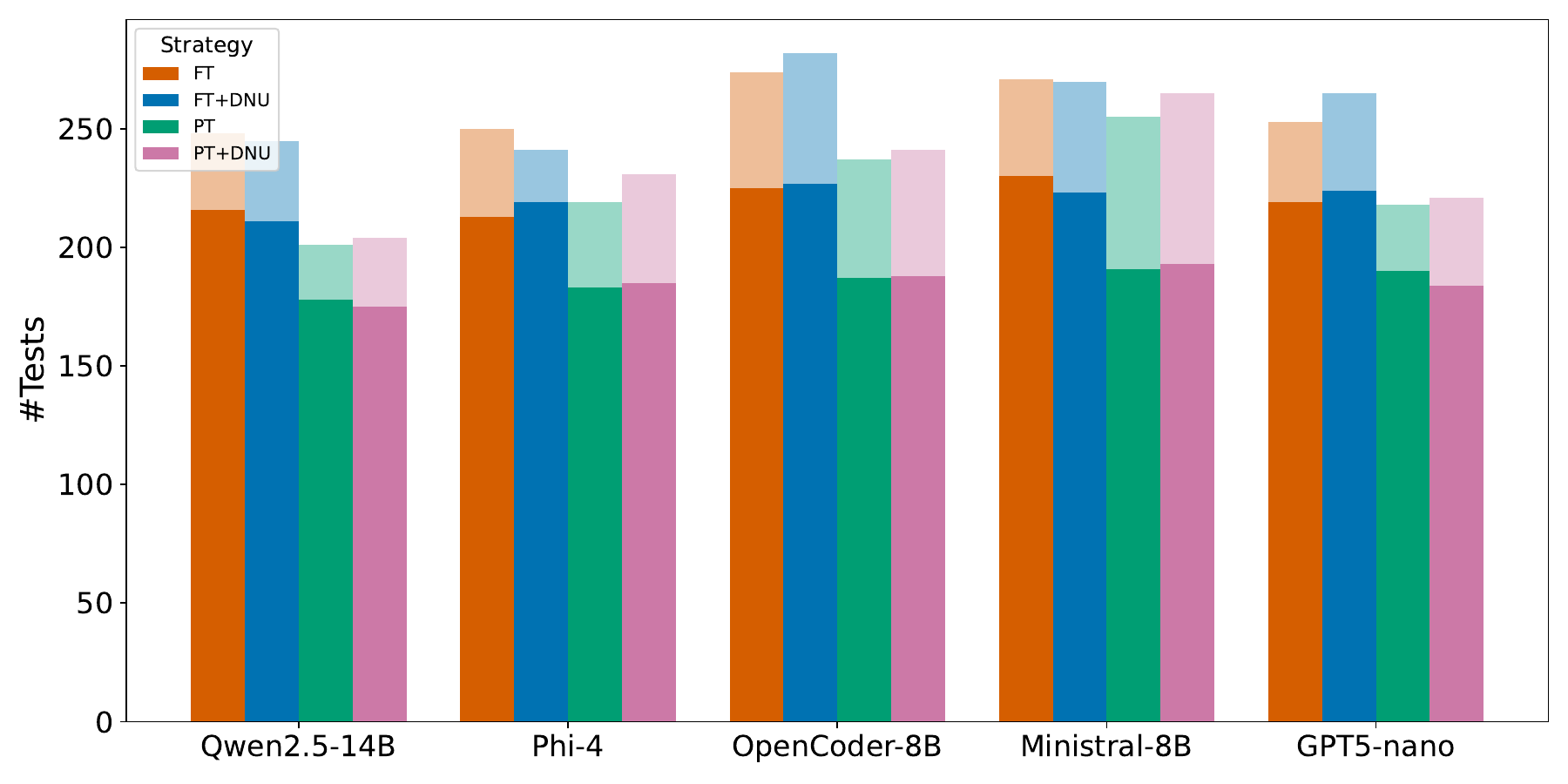}
\caption{New passing (solid) and failing (transparent) tests across models and prompting strategies compared to the baseline.}
\label{fig:new_pass_fail}
\end{figure}

Finally, we extend our analysis beyond pass@$k$ to examine \emph{new pass} and \emph{new fail} counts, which capture the number of individual tests (out of 854) that are newly solved or newly broken compared to the baseline specification. As shown in Figure~\ref{fig:new_pass_fail}, all prompting strategies substantially expand the set of solved tests by $35\%-55\%$. With FT, models achieve between $+213$ and $+230$ new passes, while FT+DNU reaches a comparable range of $+211$ to $+227$, again confirming that models benefit from visible tests even under explicit restriction. Partial test strategies (PT and PT+DNU) yield smaller but still consistent gains, adding $+175$ to $+193$ new passes across models. At the same time, all strategies introduce regressions in the form of new failing tests: FT results in $32$–$49$ new fails, while PT+DNU produces $29$–$72$. Nevertheless, these regressions are markedly smaller in scale than the corresponding gains, showing that the benefits of test exposure far outweigh the drawbacks. Overall, embedding tests substantially increases correctness by enabling models to solve many additional cases, though it introduces some new errors.

The observed performance gains raise key questions about how LLMs exploit information embedded in prompts and how this shapes their generated code. Improvements in pass@$k$ and \emph{new pass} suggest that models leverage signals from the provided tests, even when explicitly instructed not to use them. Importantly, this behavior extends beyond the full-test setting: partial and restricted variants exhibit the same tendency, showing that even limited test visibility provides exploitable cues. These findings underscore the need to better understand model behavior when balancing competing training and alignment objectives, namely, generating accurate code by exploiting all available signals while adhering to implicit or explicit restrictions on test case usage. To this end, in \emph{RQ2}, we examine whether different prompting strategies produce generations that diverge substantially from the reference solutions, and we analyze how these divergences manifest in the structure of the generated code.

\begin{leftbar}
\textbf{Answer to RQ1.} 
Embedding test cases in prompts consistently improves LLM performance in code generation. 
Full test visibility yields the largest gains, partial tests provide moderate improvements, and explicit restrictions do not significantly reduce these benefits.
However, these gains are accompanied by a smaller number of new failing tests, indicating that test exposure also introduces some regressions.
\end{leftbar}

\subsection{RQ2. Impact on the Generated Code}


The next step is to examine how the solutions generated by our five prompting strategies differ from the reference solution provided in the benchmark. Table~\ref{tab:codebleu_loc} presents the LOC and CodeBLEU results relative to the reference, comparing models and prompting strategies.


\begin{table}[ht]
\centering
\caption{CodeBleu and LOC results for different prompting strategies across models.}
\label{tab:codebleu_loc}
\resizebox{.9\textwidth}{!}{%
\begin{tabular}{lccccc}
\hline
\textbf{Strategy} & \textbf{Qwen2.5-14B} & \textbf{Phi-4} & \textbf{OpenCoder-8B} & \textbf{Ministral-8B} & \textbf{GPT5-nano} \\
\hline
\multicolumn{6}{c}{\textit{CodeBLEU}} \\
\hline
Baseline                 & 47.35 & 46.69 & 49.55 & 45.20 & 46.38 \\
FT               & 49.27 \textbf{(+4.05\%)} & 48.95  \textbf{(+4.84\%)} & 51.10 \textbf{(+3.13\%)} & 48.46 \textbf{(+7.21\%)} & 47.85 (\textbf{3.16\%)}\\
FT+DNU & 48.77 \textbf{(+2.99\%)} & 48.94 \textbf{(+4.82\%)} & 50.98 \textbf{(+2.89\%)} & 48.33 \textbf{(+6.92\%)} & 47.87 (\textbf{3.22\%}) \\
PT          & 48.49 \textbf{(+2.41\%)} & 48.33 \textbf{(+3.51\%)} & 50.86 \textbf{(+2.64\%)} & 47.52 \textbf{(+5.13\%)} & 47.72 (\textbf{2.89\%}) \\
PT+DNU & 48.48 \textbf{(+2.39\%)} & 48.04 \textbf{(+2.89\%)} & 50.88 \textbf{(+2.68\%)} & 47.43 \textbf{(+4.93\%)} & 47.73 (\textbf{2.92\%})\\
\hline
\multicolumn{6}{c}{\textit{LOC} ($LOC_{reference}$ = 19.93)}\\
\hline
Baseline                 & 28.69 & 29.71 & 26.76 & 25.72 & 34.94 \\
FT               & 28.03 \textbf{(-2.31\%)} & 29.96 \textbf{(+0.84\%)} & 26.13 \textbf{(-2.35\%)} & 28.41 \textbf{(+10.45\%)} & 36.66 (\textbf{4.93\%})\\
FT+DNU & 28.03 \textbf{(-2.28\%)} & 28.93 \textbf{(-2.62\%)} & 26.09 \textbf{(-2.47\%)} & 26.67 \textbf{(+3.68\%)} & 36.39 (\textbf{4.15\%})\\
PT          & 27.97 \textbf{(-2.50\%)} & 29.67 \textbf{(-0.14\%)} & 25.86 \textbf{(-3.33\%)} & 28.63 \textbf{(+11.29\%)} & 34.75 (\textbf{-0.53\%})\\
PT+DNU & 27.65 \textbf{(-3.63\%)} & 28.79 \textbf{(-3.09\%)} & 25.77 \textbf{(-3.69\%)} & 26.37 \textbf{(+2.52\%)} & 35.18 (\textbf{0.71\%})\\
\hline
\end{tabular}
}
\end{table}


In terms of \emph{CodeBLEU}, similarity with the reference solution increases slightly but consistently when test cases are embedded in the prompt, regardless of whether test visibility is full or partial, or of the associated restrictions. Across all models, full-test prompting (FT) yields relative improvements of about $3$–$7\%$ over the baseline, with \textit{Phi-4} and \textit{Ministral-8B} showing the largest gains ($+4.8\%$ and $+7.2\%$, respectively). Even partial test prompting provides clear benefits, indicating that models leverage the visible subset of tests to align more closely with the reference solution. Explicit restrictions have negligible impact: FT+DNU and PT+DNU achieve nearly the same improvements as their unrestricted counterparts, echoing the findings of \emph{RQ1} that models continue to exploit test visibility despite explicit prohibitions.  

In terms of \emph{LOC}, the analysis shows that generated solutions are generally longer than the reference solutions ($LOC_{reference}=19.93$). Baseline outputs typically exceed the reference by $30$–$70\%$, reflecting a tendency of LLMs to over-generate. In some cases, however, LOC moves closer to the reference when tests are visible. For instance, \textit{OpenCoder-8B} reduces its average LOC by $-2.35\%$ with FT compared to its baseline, while \textit{Ministral-8B} remains the most verbose model, with increases exceeding $+10\%$. These contrasting outcomes point to model-specific adaptation behaviors: some models respond to test exposure by becoming more concise, while others expand their outputs, often by inserting additional checks or auxiliary functions.


The combined analysis of \emph{CodeBLEU} and \emph{LOC} indicates that test exposure consistently improves structural similarity with the reference solutions, as reflected in the CodeBLEU gains across models, but its effect on LOC diverges. \textit{Qwen2.5} and \textit{OpenCoder-8B} become slightly more concise, \textit{Phi-4} fluctuates around its baseline, while \textit{Ministral-8B} produces substantially longer outputs and \textit{GPT5-nano} shows modest length increases. These divergent adaptations suggest that models pursue alignment through different strategies—some simplify code to tighten similarity, while others expand it with additional logic, often at the cost of verbosity.

Overall, test exposure consistently steers generations closer to the reference in terms of CodeBLEU, yet the varying LOC deviations reveal heterogeneous adaptation behaviors across models. Crucially, the persistence of improvements under DNU conditions shows that explicit restrictions do not negate the influence of test cases; rather, models appear to internalize these signals indirectly, reshaping both code length and structure. These findings extend the insights from \emph{RQ1}: while the different prompt strategies were shown to improve correctness (pass@$k$), the present analysis demonstrates that it also systematically alters the form of the generated code.

\begin{leftbar}
\textbf{Answer to RQ2.} 
The different prompting strategies yield systematically different generations compared to the BigCodeBench reference solutions. 
Test exposure consistently improves CodeBLEU scores and affects code length, with models exhibiting distinct adaptation behaviors. 
Crucially, these increases persist even under explicit restrictions, indicating that models continue to exploit signals embedded in the visible tests. 
Prompting strategies not only lead to higher correctness but also cause models to reshape the structure and style of their generated code.
\end{leftbar}

\subsection{RQ3. Code Variations between Prompting Strategies}


In RQ2, we examined how test visibility affects the similarity of generated solutions to the reference solutions in BigCodeBench. In RQ3, the focus shifts to how the generated code evolves compared to the baseline condition without tests. Here, CodeBLEU is recalculated relative to the baseline rather than the reference, complemented by churn-based metrics (added and removed lines) to capture structural rewrites. This allows us to assess the extent to which test exposure, full, partial, or restricted, induces meaningful changes in the generated code. To determine whether these observed evolutions are statistically robust, we apply the Wilcoxon signed-rank test, a non-parametric method for paired data, with False Discovery Rate (FDR) correction for multiple comparisons.

Table~\ref{tab:churn_metrics} reports the mean number of added and deleted lines when generations under different prompting strategies are compared against the baseline. The largest churn occurs when moving from the baseline to PT and FT, confirming that access to visible tests triggers substantial structural rewrites. For example, \textit{Phi-4} and \textit{Ministral-8B} introduce more than 28 new lines on average, accompanied by over ten deletions, evidence that test exposure drives extensive restructuring rather than minor edits. In contrast, \textit{OpenCoder-8B} and \textit{Qwen2.5} exhibit more moderate churn (7–14 added lines), indicating lighter but still systematic adaptations. 


\begin{table}[!htbp]
\centering
\caption{Churn metrics (mean lines added [+], deleted [–], and total) for version comparisons across models.}
\label{tab:churn_metrics}

\resizebox{1\textwidth}{!}{%
\begin{tabular}{ll ccc ccc ccc ccc ccc}
\toprule
 & & \multicolumn{3}{c}{\textbf{Ministral-8B}} & \multicolumn{3}{c}{\textbf{OpenCoder-8B}} & \multicolumn{3}{c}{\textbf{Phi-4}} & \multicolumn{3}{c}{\textbf{Qwen2.5-14B}} & \multicolumn{3}{c}{\textbf{GPT5-nano}} \\
\cmidrule(lr){3-5} \cmidrule(lr){6-8} \cmidrule(lr){9-11} \cmidrule(lr){12-14} \cmidrule(lr){15-17}
\textbf{Version 1} & \textbf{Version 2} & + & – & Total & + & – & Total & + & – & Total & + & – & Total & + & – & Total \\
\midrule
Baseline & PT       & \textbf{28.31} & 12.86 & 41.17 & 7.91 & 8.07 & 15.98 & \textbf{29.23} & 9.63 & 38.86 & 14.47 & 8.16 & 22.63 & 13.57 & 12.97 & 26.54 \\
Baseline & PT+DNU   & 7.93 & 8.38 & 16.31 & 7.20 & 8.05 & 15.25 & 7.52 & 8.65 & 16.17 & 6.55 & 7.67 & 14.22 & 14.05 & 12.97 & 27.02 \\
Baseline & FT       & \textbf{28.49} & 12.85 & 41.34 & 8.39 & 8.45 & 16.84 & \textbf{32.72} & 10.82 & 43.54 & 13.93 & 8.59 & 22.52 & 15.56 & 13.53 & 29.09 \\
Baseline & FT+DNU   & 9.01 & 9.26 & 18.27 & 7.66 & 8.22 & 15.88 & 8.98 & 9.91 & 18.89 & 7.64 & 8.30 & 15.94 & 15.14 & 13.53 & 28.67 \\
PT & PT+DNU         & 8.87 & \textbf{24.76} & 33.63 & 1.97 & 2.66 & 4.63 & 4.34 & \textbf{25.07} & 29.41 & 2.85 & 10.29 & 13.14 & 11.89 & 11.41 & 23.30 \\
FT & FT+DNU         & \textbf{24.91} & 8.78 & 33.69 & 2.82 & 1.88 & 4.70 & \textbf{22.91} & 10.06 & 32.97 & 7.35 & 5.69 & 13.04 & 11.83 & 11.21 & 23.04 \\
\bottomrule
\end{tabular}
}
\end{table}




Restricted conditions (PT+DNU and FT+DNU) yield smaller churn relative to their unrestricted counterparts. When moving from the baseline to PT+DNU or FT+DNU, all models add and remove fewer lines (typically under 10), suggesting that explicit restrictions partially dampen the extent of structural change. Yet, the fact that churn values remain above zero indicates that models continue to adapt their generations even under DNU instructions. This aligns with earlier findings in \emph{RQ1} and \emph{RQ2} that restrictions do not completely prevent models from exploiting test-related cues.  

Direct comparisons between restricted and unrestricted variants (PT vs. PT+DNU, FT vs. FT+DNU) reveal how strongly models react to explicit “Do Not Use” (DNU) instructions. \textit{Phi-4} and \textit{Ministral-8B} show exceptionally high churn in these contrasts (over 20 lines added or deleted), indicating that their generations diverge substantially when tests are visible but flagged as unusable, evidence that they actively adjust to the restriction and thus display better instruction-following. By contrast, \textit{Qwen2.5} shows smaller yet consistent differences, while \textit{OpenCoder-8B} exhibits only limited churn, suggesting weaker adherence to the DNU restriction. \textit{GPT5-nano} remains largely invariant across conditions, implying that its generations are minimally influenced by explicit prohibitions. 

Overall, the churn analysis shows that test exposure consistently induces structural rewrites, while explicit restrictions reshape rather than eliminate these adaptations. The degree of change varies markedly across models. \textit{Ministral-8B} and \textit{Phi-4} are the most sensitive to prompt conditions, with substantial churn reflecting 
strong test exploitation and active adjustment under DNU instructions. \textit{Qwen2.5} displays intermediate sensitivity, adapting in a more moderate but consistent way. \textit{OpenCoder-8B} shows limited 
divergence, suggesting weaker responsiveness to restrictions, whereas \textit{GPT5-nano} remains 
stable across all conditions, with minimal evidence of adaptation.

To evaluate whether the observed behavioral differences are statistically robust, we first tested the distributions for normality using the Kolmogorov–Smirnov (KS) test \cite{berger2014kolmogorov}. Since several samples violated the normality assumption, we relied on non-parametric methods. Specifically, we applied the Wilcoxon signed-rank test \cite{woolson2007wilcoxon} for paired comparisons across prompting strategies. This test does not assume normality and is well suited for detecting systematic differences in paired samples. To control for multiple comparisons, we adjusted the resulting $p$-values using the Benjamini–Hochberg False Discovery Rate (FDR) procedure \cite{benjamini1995controlling}.

Table~\ref{tab:wilcoxon_codebleu_loc} summarizes the differences in CodeBLEU and churn between restricted and unrestricted variants (PT vs. PT+DNU, FT vs. FT+DNU) relative to the baseline, along with their significance levels after FDR correction.


\begin{table*}[ht]
\centering
\caption{Mean \emph{CodeBLEU} scores and \emph{total churn} relative to the baseline, along with mean differences for \emph{PT vs PT+DNU} and \emph{FT vs FT+DNU}. Statistical significance of the differences is assessed using the Wilcoxon signed-rank test. Stars indicate significance: $^{*}$: $p < 0.05$, $^{**}$: $p < 0.01$.}
\label{tab:wilcoxon_codebleu_loc}
\begin{subtable}{1\linewidth}
\centering
\caption{Mean CodeBLEU scores relative to the baseline, and mean differences for \emph{PT vs PT+DNU} and \emph{FT vs FT+DNU}, with statistical significance assessed using the Wilcoxon signed-rank test.}
\begin{tabular}{lccc|ccc}
\toprule
 & \multicolumn{3}{c|}{PT vs PT+DNU} & \multicolumn{3}{c}{FT vs FT+DNU} \\
Model & PT & PT+DNU & Diff & FT & FT+DNU & Diff \\
\midrule
Ministral      & 53.63 & 68.65 & \textbf{-15.02}$^{**}$ & 51.18 & 64.97 & \textbf{-13.79}$^{**}$ \\
OpenCoder      & 67.28 & 68.19 & -0.91                  & 65.81 & 66.90 & \textbf{-1.08}$^{*}$   \\
Phi-4              & 52.61 & 68.95 & \textbf{-16.34}$^{**}$ & 50.26 & 64.33 & \textbf{-14.07}$^{**}$ \\
Qwen2.5\_Coder & 64.01 & 71.51 & \textbf{-7.50}$^{**}$  & 62.80 & 68.31 & \textbf{-5.51}$^{**}$  \\
openai-gpt5-nano   & 58.57 & 58.22 & 0.34                   & 55.46 & 56.25 & -0.79                  \\
\bottomrule
\end{tabular}
\end{subtable}%
\hfill
\vspace{5pt}
\begin{subtable}{1\linewidth}
\centering
\caption{Mean total churn relative to the baseline, and mean differences for \emph{PT vs PT+DNU} and \emph{FT vs FT+DNU}, with statistical significance assessed using the Wilcoxon signed-rank test.}
\begin{tabular}{lccc|ccc}
\toprule
 & \multicolumn{3}{c|}{PT vs PT+DNU} & \multicolumn{3}{c}{FT vs FT+DNU} \\
Model & PT & PT+DNU & Diff & FT & FT+DNU & Diff \\
\midrule
Ministral      & 41.18 & 16.31 & \textbf{24.86}$^{**}$ & 41.34 & 18.26 & \textbf{23.08}$^{**}$ \\
OpenCoder     & 15.99 & 15.25 & 0.74                  & 16.84 & 15.88 & \textbf{0.97}$^{**}$  \\
Phi-4              & 38.86 & 16.17 & \textbf{22.69}$^{**}$ & 39.28 & 14.64 & \textbf{24.64}$^{**}$ \\
Qwen2.5\_Coder & 26.18 & 17.77 & \textbf{8.41}$^{**}$  & 25.86 & 19.28 & \textbf{6.58}$^{**}$  \\
openai-gpt5-nano   & 16.37 & 16.83 & -0.47                 & 16.32 & 15.89 & 0.43                  \\
\bottomrule
\end{tabular}
\end{subtable}%
\end{table*}

For \textit{Ministral-8B}, \textit{Phi-4}, and \textit{Qwen2.5}, nearly all pairwise comparisons reach high statistical significance ($p < 0.01$) for both CodeBLEU and churn, confirming substantial behavioral shifts when explicit restrictions are introduced. \textit{OpenCoder-8B} shows significant differences in some cases—most notably FT vs. FT+DNU for both CodeBLEU and churn—but remains stable in others. In contrast, \textit{GPT5-nano} exhibits no statistically significant differences across any of the tested pairs. These results demonstrate that most LLMs adapt their generations in a statistically robust way under varying visibility and restriction conditions, though the extent of adaptation varies considerably across models.

\begin{leftbar}
\textbf{Answer to RQ3.} 
Churn analysis shows that exposing models to tests, whether partially or fully, induces substantial structural rewrites, often involving dozens of added or removed lines relative to baseline generations. These effects are most pronounced for \textit{Phi-4} and \textit{Ministral-8B}, while \textit{Qwen2.5}, \textit{OpenCoder-8B}, and \textit{GPT5-nano} exhibit more moderate changes. The Wilcoxon tests confirm that many of these differences are statistically significant for both CodeBLEU and churn, particularly among the open-source models. Taken together, these findings indicate that test visibility systematically reshapes model outputs under both restricted and unrestricted instructions, and that the resulting behavioral adaptations are consistent and statistically robust rather than incidental.
\end{leftbar}




\subsection{RQ4. Adaptation Strategies}
The analyses for the first three research questions revealed consistent behavioral variations when tests were visible, both with and without explicit restrictions. In RQ4, we turn to a qualitative examination of the code generated under different prompting strategies to better understand the nature of these variations across models, prompting strategies, and tasks. From this analysis, we identify four distinct adaptation strategies that LLMs employ when exposed to test suites, offering valuable context for interpreting the quantitative results. For each strategy, we illustrate representative tasks, prompts, and generated outputs.
The qualitative study was conducted manually on more than 30 of the 148 BigCodeBench (Hard) tasks, chosen because they exhibited substantial metric differences across prompting strategies. Throughout this section, we refer to tasks by their IDs in the BigCodeBench dataset (which is why most IDs exceed 148).


\subsubsection{Code Refinement Using Test Signals}\label{sec:strategy_refinement}

\begin{figure}
\includegraphics[width=.9\linewidth]{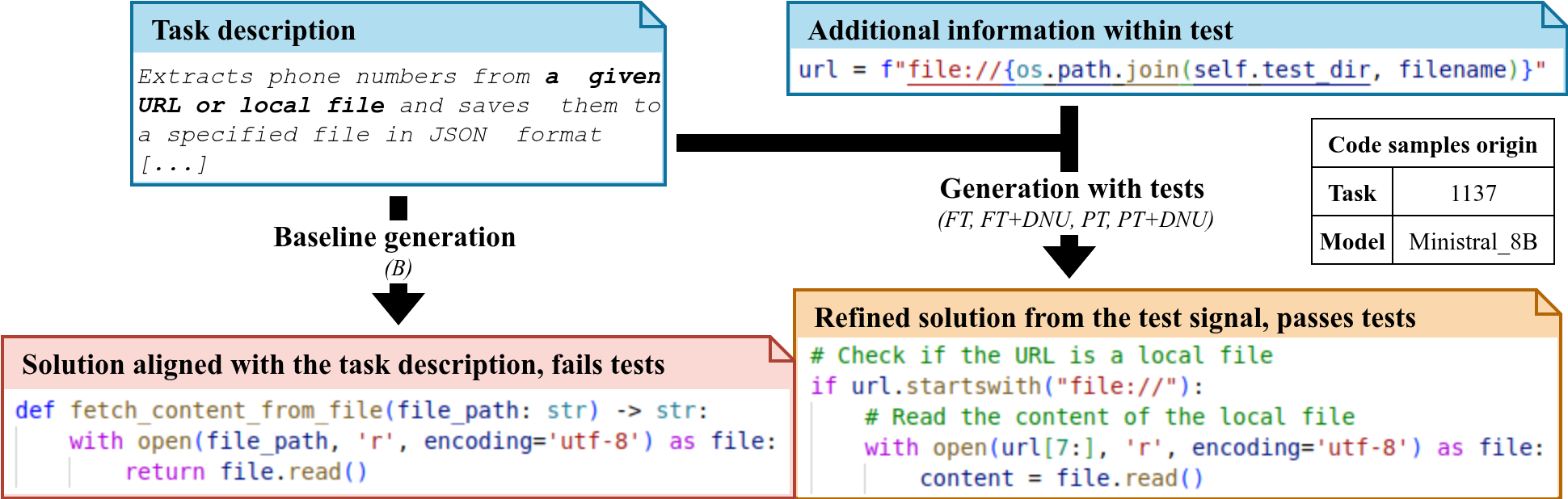}
\caption{Code Refinement Using Test Signals: the solution is adapted to better fit the test suite}
\label{fig:1137_solution_refinement}
\end{figure}

When exposed to a test suite, the studied LLMs consistently adapted their code to better align with the behavior implied by the tests. Rather than treating the problem description as the sole specification, the models appeared to interpret the tests as authoritative guidance and refined their outputs accordingly.
For instance, in task 1137 (see Figure~\ref{fig:1137_solution_refinement}), the description requested extracting data from ``\textit{a given URL or local file}.'' Without access to tests, all LLMs assumed that local files would be provided as directory paths (e.g., \verb|/home/user/path_to_file|). When exposed to tests, however, the models converged on a different interpretation: treating local files as file URLs (e.g., \verb|file://path_to_file|) and adjusting their code to handle only this case. Similar refinements were observed in numerous other tasks. For example, in task 15, models modified their output to produce the exact error message expected by the tests, and in task 765, they adapted their handling of dictionary parameters to satisfy test requirements.
This refinement strategy was frequently adopted even under explicit instructions not to use the tests, suggesting that alignment with the pretraining objective of maximizing accuracy often outweighed prompt-based constraints. \emph{Ministral} and \emph{Phi-4}, in particular, demonstrated strong tendencies to enforce such refinements despite restrictions.

Overall, this was the most prevalent adaptation strategy. Its dominance is consistent with the notable improvements observed in pass@1 and pass@5 metrics under test exposure. LLMs systematically exploit test signals to refine their code, even when explicitly instructed not to.



\subsubsection{Test Hard Coding}

In extreme cases, attempts to exploit test signals too closely can cause elements from the tests to leak directly into the generated solution, often compromising consistency with the task description or even with other tests in the suite.

\begin{figure}
\includegraphics[width=.8\linewidth]{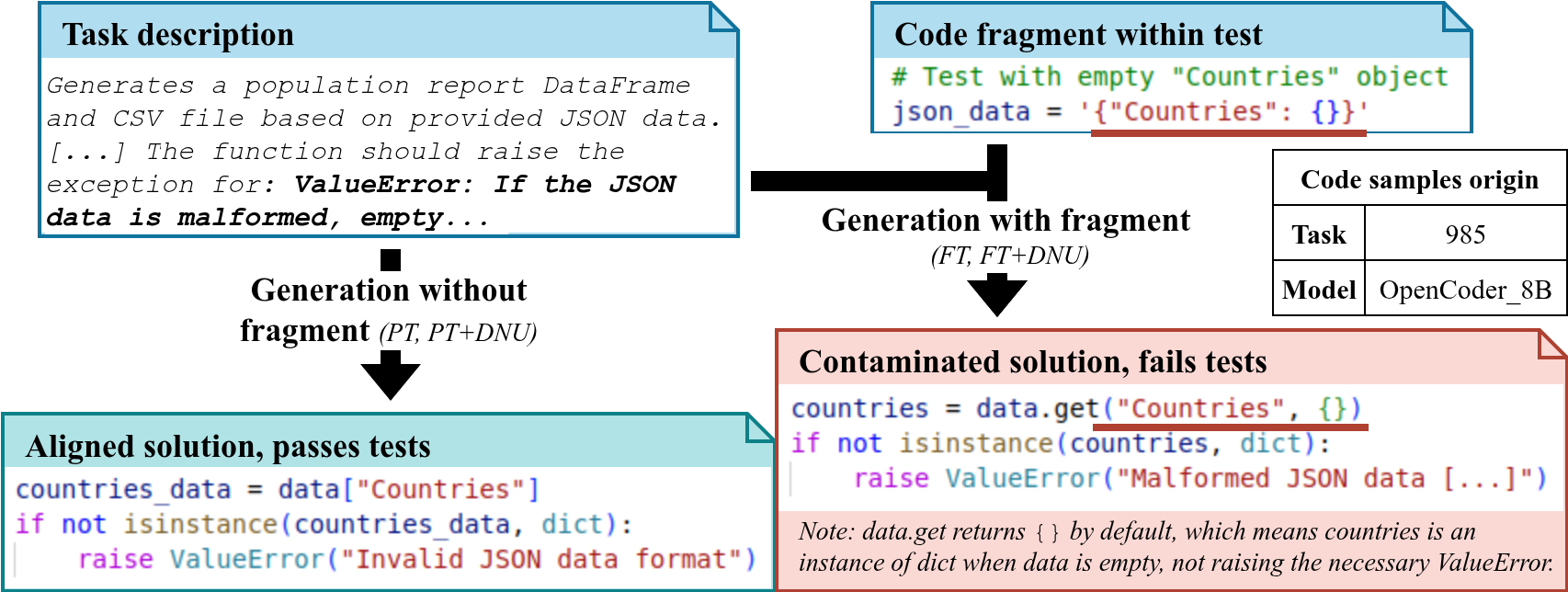}
\caption{Test Hard Coding: code from the test suite incoherently leaks into the generated code }
\label{fig:985_hard_coding}
\end{figure}

For example, in task 985 (see Figure~\ref{fig:985_hard_coding}), the task requires processing JSON data containing a dictionary and raising an exception if the data is missing. When only partial tests are available, \textit{OpenCoder} produces correct code that passes all tests. However, when exposed to the full test suite, the solution fails. The failure arises from initializing the dictionary as \begin{small}\texttt{countries = data.get("Countries", \{\})}\end{small}'', which returns an empty dictionary \begin{small}\verb|{}|\end{small}'' when the data is absent, preventing the required exception from being raised. This issue is directly linked to the test suite itself: one of the final tests explicitly initializes an empty dictionary with ``\begin{small}\verb|json_data = '{"Countries": {}}'|\end{small}'', reproducing the structure generated by the failing code. In this case, the test signals override alignment with the original task specification, illustrating how over-reliance on test cues can introduce errors. Notably, this behavior occurs even under the Do Not Use variants of both Full Tests and Partial Tests instructions.

A similar phenomenon appears in task 310, where \textit{Qwen25\_Coder\_14B} incorrectly applies the \verb|zip| function by imitating its use from a test. Again, this error manifests only under Full Tests conditions (i.e., the FT/FT+DNU code fails, whereas the PT/PT+DNU code passes), showing that excessive adaptation to test signals can lead to task-specific mistakes.


\subsubsection{Quick \& Dirty Programming}


Access to the test suite can lead LLMs to focus narrowly on passing the tests, often at the expense of broader code quality attributes such as reusability, maintainability, or generalizability, a behavior we characterize as ``quick and dirty programming”.

\begin{figure}
\includegraphics[width=0.75\linewidth]{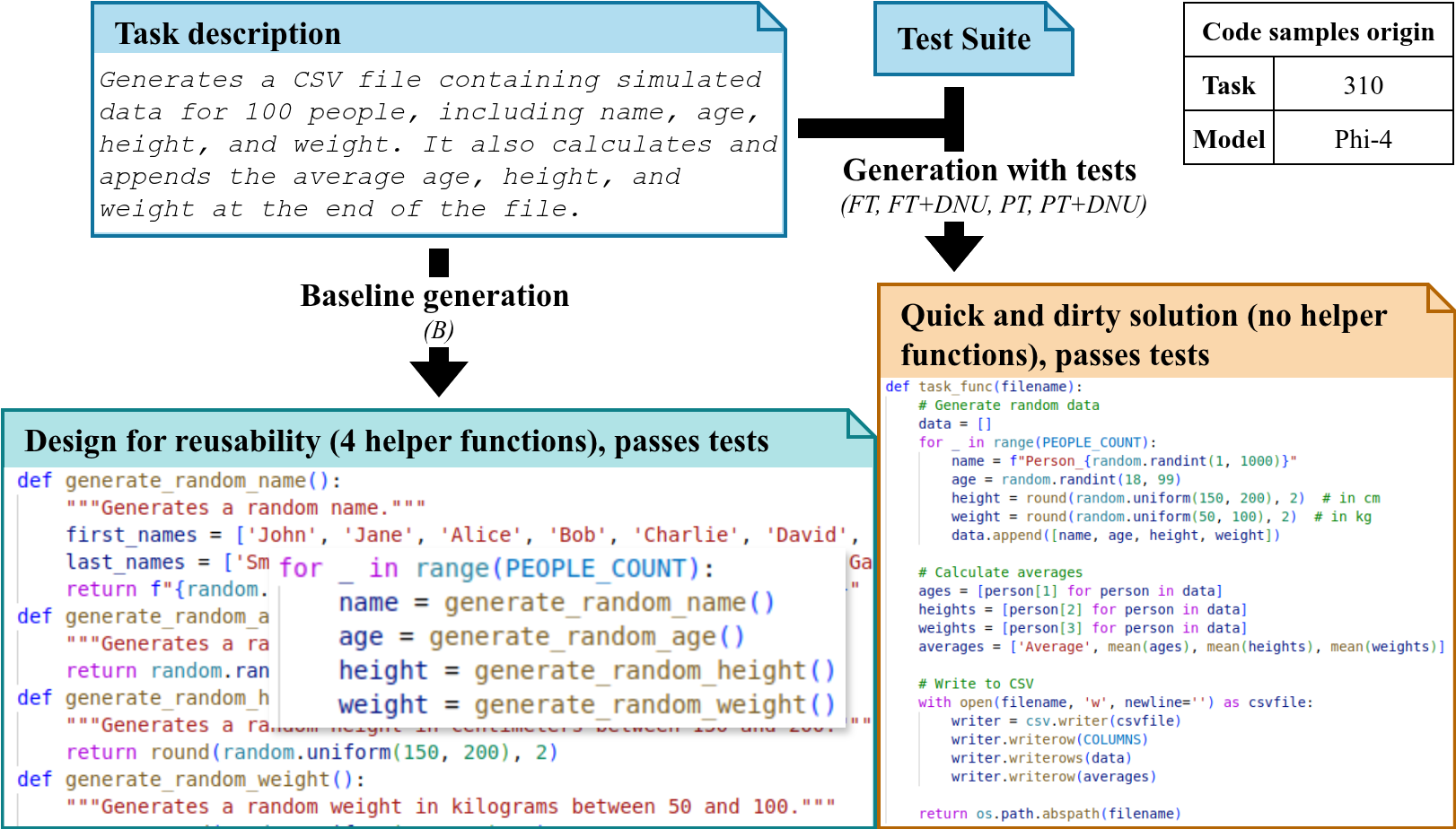}
\caption{Quick \& Dirty Programming: generates passing code, but disregards other properties.}
\label{fig:310_quick_dirty}
\end{figure}

In task 310 (Figure~\ref{fig:310_quick_dirty}), for example, the LLM must generate a CSV. For \textit{Phi-4}, the task is solved correctly under all prompting strategies, and the tests are consistently passed. Yet the structure of the solution differs markedly. Without access to the test suite, the model decomposes the program into four subfunctions, yielding a modular and reusable design. When tests are visible (across all other prompting strategies), the solution is simplified to a flat structure with at most one helper function, reflecting a clear decline in generalizability and reusability.

We found many examples of this pattern. In task 1137 (discussed in Section~\ref{sec:strategy_refinement}), the baseline generation produced helper functions and test-exposed conditions yielded monolithic code. Across multiple LLMs and tasks, we observe the same adaptation strategy: models prioritize immediate test-passing functionality over sustainable design, trading long-term robustness and readability for short-term effectiveness. While this ensures correctness with respect to tests, it risks producing brittle solutions in real-world settings where code quality and maintainability are essential.

In contrast to the other two adaptation strategies, refinement, where models iteratively improve the structure of their solutions, and test code reuse, where they incorporate elements of the provided tests as templates for generation, quick and dirty programming represents a regression in code quality. This strategy shows how the presence of tests can push models toward expedient but fragile solutions, highlighting a critical tension between alignment signal–driven compliance and the broader goals of practical software engineering.

\subsubsection{No Adaptation}


We also identified a fourth category where LLMs showed little or no adaptation to the test suite. In these cases, the generations aligned with the instruction not to use tests, and the code exhibited no clear evidence of test influence. Minor variations did appear (e.g., variable names or comments), but the underlying solutions remained essentially unchanged.

\begin{figure}
\includegraphics[width=0.7\linewidth]{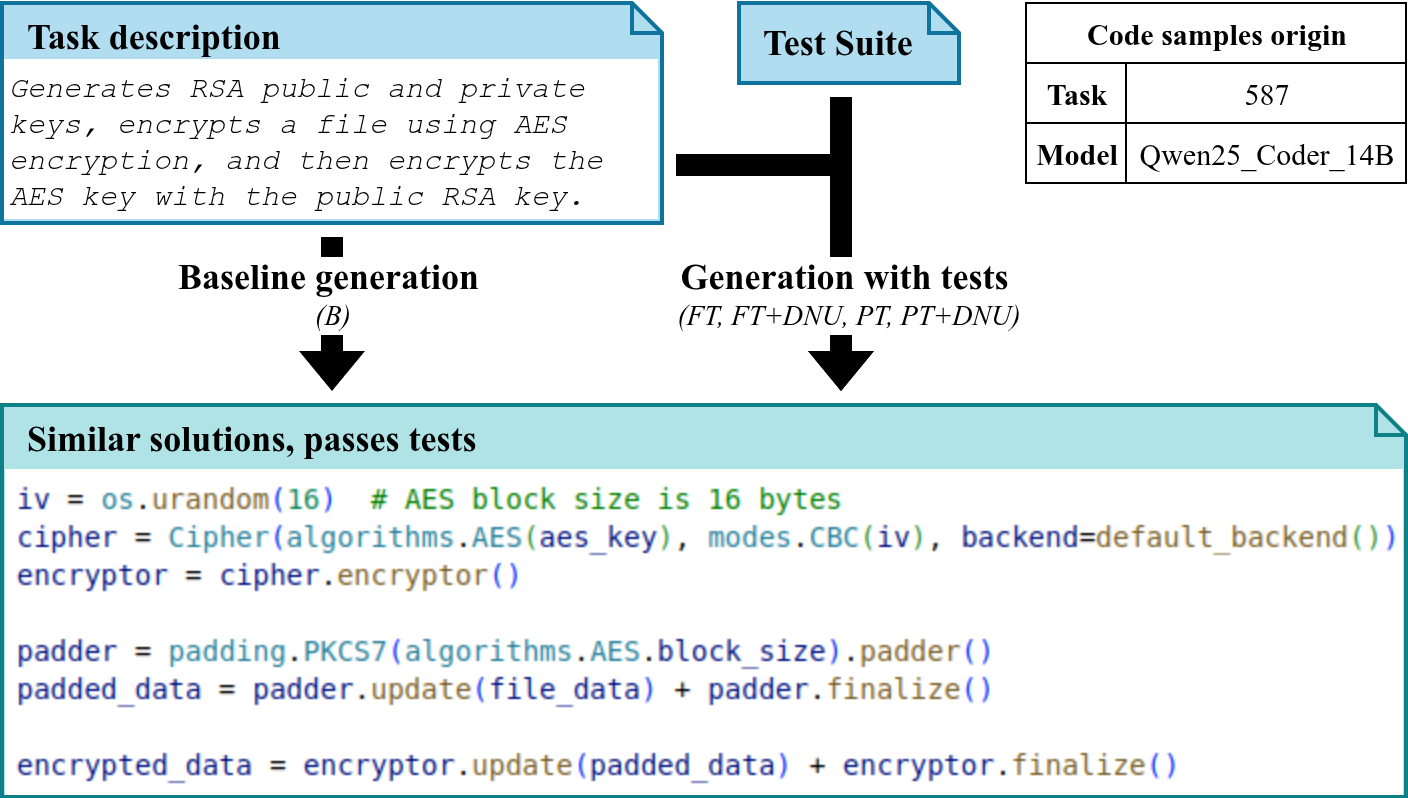}
\caption{No Adaptation: generated code isn't alterned after exposition to the test suite.}
\label{fig:587_no_adaptation}
\end{figure}

For instance, in task 587 (Figure~\ref{fig:587_no_adaptation}), which requires implementing a basic encryption algorithm, all models produced functionally equivalent code across prompting strategies. The outputs passed the tests consistently, with differences limited to superficial edits.

This pattern was especially common in simple tasks, such as simulating dice rolls (task 897) or generating all subsets of a set (task 928), where models generated nearly identical correct solutions regardless of test visibility. The likely explanation is that such tasks are straightforward enough for the models to be confident in their baseline solutions, making additional signals unnecessary.

By contrast, in more challenging tasks such as building a web server (task 1040), all models failed to solve the problem and their outputs diverged widely across prompting strategies—reflecting higher uncertainty and a stronger reliance on external signals like unit tests or explicit instructions.

\begin{leftbar}
\textbf{Answer to RQ4.} 
We identified four distinct adaptation strategies that LLMs employ under varying levels of test visibility and restriction. The most common is refinement, where access to the test suite helps models adjust and improve their solutions to better match the task specification. At the opposite extreme, some models engage in test code reuse, directly hard-coding outputs or copying test fragments, which often results in brittle or failing solutions. A third strategy, quick and dirty programming, arises when models prioritize passing the tests over producing reusable or generalizable code, trading long-term quality for short-term functionality. Finally, in simpler tasks, models frequently exhibited no adaptation, generating confident solutions that were essentially unaffected by the presence or absence of test signals.
\end{leftbar}

\section{Threats to Validity and Discussion}
\label{sec:discussion}

In this section, we discuss potential threats to the validity of our study and reflect on the broader implications of our results. While limitations are inevitable in empirical work, we view them as opportunities to guide future investigations 
on LLM behavior in code generation.

\textit{Prompt design.} Our study did not aim to conduct a comprehensive prompt-engineering exploration. Other prompt formulations beyond those we tested may lead to different outcomes. Nevertheless, the prompting strategies we selected capture meaningful variations and reveal consistent behavioral trends, suggesting that the phenomena we observed are not artifacts of a specific design but reflect more general properties of LLMs.

\textit{Metrics.} We relied on established quantitative metrics, such as \texttt{pass@k}, CodeBLEU, and code churn, to capture correctness and structural differences. These measures provided useful insights but cannot fully characterize more subtle behaviors, including opportunistic or deceptive strategies. At present, there are no widely accepted metrics for detecting deception in code generation. Existing work has relied on hand-curated evaluations \cite{wu2025opendeception} or LLM-based judgments \cite{ferreira2025truthful}. This highlights a promising avenue for future research: the design of dedicated evaluation frameworks for identifying misalignment and deceptive tendencies in code generation models.

\textit{Manual analysis of adaptation strategies.} The four adaptation strategies we identified emerged from manual inspection of a subset of tasks. This process is inherently subjective and covers only part of the dataset, which limits claims about prevalence. Still, these strategies appeared repeatedly across tasks and models, lending confidence that they capture robust patterns of adaptation. Future work could expand this analysis with automated or hybrid human–AI coding methods to quantify the frequency and conditions under which each strategy arises.

\textit{Benchmark scope.} Our experiments were conducted on BigCodeBench, whose unit tests are often more precise and restrictive than natural language task descriptions. This makes them unusually strong alignment signals. Different benchmarks with looser test specifications might elicit other behaviors, such as overfitting to weak test suites, an effect we did not observe here. Moreover, BigCodeBench focuses exclusively on Python, leaving open questions about generalizability to other languages with different idioms, libraries, and ecosystems.

\textit{Model scope} We focused on medium-to-large models representative of today’s open-source landscape. Larger frontier models and those prompted with advanced reasoning strategies (e.g., chain-of-thought) may exhibit different dynamics. Prior work suggests that scale and reasoning-oriented prompting can strengthen alignment but introduce more sophisticated deceptive behaviors \cite{anthropic2025claude4}. Investigating these dimensions remains an important direction for follow-up studies.

\textit{Dataset quality.} Finally, we identified some quality issues in BigCodeBench itself. Certain tasks contained overly specific requirements (e.g., exact error messages), incorrect reference implementations, or flawed test suites (e.g., misformatted numbers in task 865, input substitution errors in task 945). Although these cases represent a small fraction of the dataset, they illustrate the challenges of constructing and maintaining large-scale, high-quality benchmarks for code generation.

\textit{Implications.} Despite these limitations, our findings yield several important insights. First, explicit restrictions not to use test cases are frequently disregarded, as models continue to exploit visible signals to improve their outputs. This underscores the difficulty of enforcing alignment when external signals strongly correlate with the training objective of predictive accuracy. Second, test exposure can shift models toward undesirable behaviors, such as code leakage or “quick-and-dirty” programming, where passing tests takes precedence over producing reusable and maintainable solutions. These observations are particularly salient in the context of agentic models that have access to all project artifacts. 
In practice, prompts may inadvertently expose logs, outdated traces, partial test cases, or intermediate outputs from earlier pipeline stages. Our experimental setup isolates this phenomenon by using unit tests as a controlled stand-in for such unintended contextual signals.
Such models could exploit multiple sources of information simultaneously, amplifying opportunistic strategies and increasing the risk of misaligned behaviors across complex development pipelines.
These insights carry broader implications for the trustworthiness and evaluation of LLMs in programming. Current benchmarks often overlook the opportunistic strategies that models employ, leaving gaps in our ability to detect and quantify misalignment. Addressing this challenge requires developing new evaluation methodologies that explicitly capture deceptive or shortcut-seeking behavior. Recent efforts, such as the OpenDeception benchmark \cite{wu2025opendeception}, represent encouraging steps in this direction. Building on such initiatives will help the community move toward more comprehensive and reliable assessments of LLMs in software engineering.

Taken together, the threats and reflections discussed here contextualize the results of RQ1–RQ4. Our analyses showed that LLMs can disregard explicit restrictions (RQ1), converge toward reference-like solutions when test signals are present (RQ2), adapt structurally in statistically significant ways (RQ3), and follow diverse adaptation strategies ranging from refinement to quick-and-dirty programming (RQ4). Acknowledging the limitations of the study allows us to situate these findings within a broader research agenda: developing stronger benchmarks, richer evaluation metrics, and more reliable alignment techniques for LLMs in software engineering, particularly in scenarios where models have comprehensive access to project artifacts.

\section{Conclusion}
\label{sec:conclusion}

This paper investigated how LLMs adapt their code generation behavior when exposed to test cases under varying prompting conditions. Leveraging the \emph{BigCodeBench-Hard} benchmark, we systematically evaluated five LLMs across correctness, similarity to reference solutions, program size, and code churn, complemented by qualitative analyses of adaptation strategies, under five prompting strategies that manipulated test visibility and imposed explicit restrictions.

Our results reveal several key insights. Test visibility substantially improves correctness, with nearly double the number of passing tests compared to the baseline, while partial exposure yields moderate gains. Explicit restrictions (\emph{Do Not Use}) are largely ineffective, as models often exploit visible signals regardless of instructions. Performance gains consistently outweigh regressions; however, models also restructure their code in response to test signals. Qualitative analysis identifies recurring adaptation strategies, including refinement using test feedback, hardcoding test solutions, and prioritizing test passing at the expense of code quality. These observations indicate that in some contexts, apparent improvements may arise from opportunistic exploitation of available signals rather than faithful adherence to alignment objectives.

Our study highlights the need for more robust benchmarks and alignment mechanisms. Enhancing task clarity and completeness could reduce overreliance on spurious cues, while alignment strategies explicitly designed to reinforce instruction-following could curb opportunistic behaviors. By combining improved benchmark design with refined alignment techniques, future work can promote more trustworthy and instruction-compliant behavior in LLM-driven code generation.


\bibliographystyle{ACM-Reference-Format}
\bibliography{bib}


\end{document}